\documentclass[useAMS,usenatbib,a4paper,referee]{mn2e}
\usepackage{graphicx}
\usepackage{amsmath}
\usepackage{txfonts}
\usepackage{mathrsfs}

\def\aap{A\&A}
\def\apjl{ApJ}

\def\apjs{ApJS}

\newcommand{\be}{\begin{equation}}
\newcommand{\ee}{\end{equation}}
\newcommand{\bea}{\begin{eqnarray}}
\newcommand{\eea}{\end{eqnarray}}

\title[Diffusive Cosmic Ray Acceleration 
at the Galactic Centre]{Diffusive Cosmic Ray Acceleration at the Galactic Centre}
\author[F. Melia and M. Fatuzzo]{F. Melia$^{1}$\thanks{Sir Thomas Lyle Fellow and Miegunyah Fellow.
E-mail: melia@physics.arizona.edu} and M. Fatuzzo$^{2}$\thanks{E-mail: fatuzzo@xavier.edu}\\
$^{1}$Department of Physics, The Applied Math Program, and Department of Astronomy, 
The University of Arizona, AZ 85721, USA \\
$^{2}$Physics Department, Xavier University, Cincinnati, OH 45207}
\begin{document}

\date{}

\pagerange{\pageref{firstpage}--\pageref{lastpage}} \pubyear{2010}

\maketitle

\label{firstpage}

\begin{abstract}
The diffuse TeV emission detected from the inner $\sim2^\circ$ of the Galaxy
appears to be strongly correlated with the distribution of molecular
gas along the Galactic ridge. Although it is not yet entirely clear whether
the origin of the TeV photons is due to hadronic or leptonic interactions,
the tight correlation of the intensity distribution with the molecular gas
strongly points to a pionic-decay process involving relativistic protons.
But the spectrum of the TeV radiation---a power law with index $\alpha\approx
-2.3$---cannot be accommodated easily with the much steeper distribution
of cosmic rays seen at Earth. In earlier work, we examined the
possible sources of these relativistic protons and concluded that
neither the supermassive black hole Sagittarius A* (identified with
the HESS source J1745-290), nor several pulsar wind nebulae dispersed
along the Galactic plane, could produce a TeV emission profile morphologically
similar to that seen by HESS. We concluded from this earlier study that
only relativistic protons accelerated throughout the inter-cloud
medium could account for the observed diffuse TeV emission from this region.
In this paper, we develop a model for diffusive proton acceleration driven
by a turbulent Alfv\'enic magnetic field present throughout
the gaseous medium. Though circumstantial,
this appears to be the first evidence that at least some cosmic rays are
accelerated diffusively within the inner $\sim300$ pc of the Galaxy.
\end{abstract}


\section{Introduction}

The earliest observations of the Galactic center with the High-Energy Stereoscopic
System (HESS) revealed the presence of several TeV point sources, including HESS
J1745--290, coincident with the supermassive black hole Sagittarius A*, and
a second type of source, such as the supernova remnant/pulsar wind nebula G0.9+0.1
(about 1$^\circ$---or roughly 144 pc at that distance---toward positive longitude $l$
of the Galactic center), distributed along the plane (Aharonian et al. 2004).
An extended period of observation since then, coupled with HESS's unprecedented
sensitivity, has provided an opportunity of subtracting these point sources from
the overall map of this region to search for the fainter, diffuse emission. We
now know that diffuse TeV radiation is produced along the Galactic-center ridge
extending over $2^\circ$, spread out roughly $0.2^\circ$ in Galactic latitude $b$.

The diffuse TeV emission is strongly correlated with the distribution of interstellar
gas (Aharonian et al. 2006). Along with the energy range accessible to HESS
($> 200$ GeV), this morphology points to the decay of neutral pions produced in
hadronic cascades as the dominant source of diffuse radiation, and is therefore
quite likely due to the scattering of relativistic cosmic rays with protons in the
ambient medium (see, e.g., Crocker et al. 2005; Ballantyne et al. 2007).
The TeV gamma rays are apparently produced within a scale height of roughly 30 pc,
similar to that of giant molecular cloud (GMC) material in this region, as traced
by its CO and CS line emission (see, e.g., Tsuboi et al. 1999). The $\sim$$10^8\;M_\odot$
of molecular gas provides a rich target of overlapping clouds for the incoming cosmic
rays.

In their paper, Wommer et al. (2008) explored several possible source(s) of energetic
hadrons at the Galactic center, and their propagation through a turbulent medium.
The survey included Sagittarius A* itself, which may be where the cosmic rays
producing HESS J1745-290 originate, and the pulsar wind nebulae dispersed along
the Galactic ridge. They also considered the possibility that the relativistic
protons may be accelerated throughout the inter-cloud medium.

The origin of these energetic hadrons is an intriguing puzzle because a simple
cosmic-ray interpretation for the diffuse TeV emission is problematic for several
reasons. For example, the observed gamma-ray spectrum does not appear to be consistent
with the distribution seen at Earth. The gamma-ray spectrum measured by HESS in the
region $|l|<0.8^\circ$ and $|b|<0.3^\circ$ (with point-source emission subtracted)
can reasonably be fit with a power law with photon index $\Gamma=2.29\pm 0.27$.
And since the spectral index
of the gamma rays tracks the spectral index of the cosmic rays themselves, the implied
cosmic ray index ($\sim$$2.3$) is much harder than that ($\sim$$2.75$) measured locally.

It is interesting to speculate that we may be seeing the first (albeit indirect)
evidence of propagation effects as the cosmic rays diffuse outwards from the center.
These energetic protons would escape from the Galaxy on an energy-dependent time
scale $t_{\rm esc}\propto E^{-\delta}$, with $\delta$$\sim$$0.4$--$0.6$ (Bhattacharjee
2000). So the injected spectrum would be flatter (by a change in index of
$\sim\delta$) than that observed here if most of the cosmic rays detected at
Earth originate at the Galactic center. The very interesting possibility
that the two distributions may still be consistent with each other deserves
further investigation as a follow-up to the work reported here.

The principal purpose of this paper is to develop a self-consistent model
for cosmic-ray acceleration within the inner $\sim2^\circ$--$4^\circ$ of the
Galaxy. Wommer et al. (2008) concluded that the conditions at the Galactic
center preclude a point-source origin for these particles. The supermassive
black hole Sagittarius A* may be responsible for producing HESS~J1745-290,
but its hadronic efflux cannot extend out beyond a latitude $\sim \pm 1^\circ$
because the protons lose their energy or scatter with the ambient medium on
much smaller scales. Other points sources, such as the known pulsar wind nebulae,
also produce a morphology with centrally peaked emission regions not consistent
with the HESS map. It appears that only cosmic rays accelerated throughout the
inter-cloud medium can produce a diffuse TeV glow consistent with the observations.

But are the particles accelerated diffusively, or do they emerge from a (presumably
large) population of sources distributed along the Galactic ridge? Wommer et al.'s
simulations indicate that the gamma-ray emissivity associated with any given object
drops by a factor $\sim 2$ within a distance of roughly $0.1^\circ$. This is effectively
the contour range of the HESS maps, so individual sources would not stand out as
long as their angular separation were less than this value. In a projected area
$\sim$$2^\circ\times 1^\circ$, this proximity would require about 50 individual
sources.  Unfortunately, the total number of TeV sources detected by HESS (many
of them presumably pulsar wind nebulae) is far smaller than this. Other possibilities
include low-mass X-ray binaries, but only $\sim$$5$ of them have been identified
in this region (Bird et al. 2007). Other classes of object with a volume
density greater than this apparently do not produce relativistic hadrons.

Our working assumption here will therefore be that the cosmic rays observed
along the Galactic-center ridge are accelerated diffusively throughout the
inter-cloud medium. In the next section, we will summarize the available data
pertaining to the gas and magnetic field distributions at the Galactic center,
and then describe a scenario in which protons may be accelerated to TeV
energies (and beyond) due to the presence of a turbulent Alfv\'enic magnetic
field. We will follow their evolution in energy space, and calculate the
spectrum of hadrons impacting the molecular gas. And from the $pp$-induced
pion decays, we will calculate the TeV spectrum for direct comparison with
the data.

\section[]{The Physical Conditions}
The large concentration (up to $\sim$$10^8\;M_\odot$) of dense molecular
gas at the Galactic center is largely confined to GMC's with a size
$\sim$$50$--$70$ pc (G\"{u}sten and Philipp 2004). These clouds appear to
be clumpy with high-density ($\sim$$10^5$ cm$^{-3}$) regions embedded
within less dense ($\sim$$10^{3.7}$ cm$^{-3}$) envelopes (e.g., Walmsley et al.
1986). The average cloud density is therefore roughly $10^4$ cm$^{-3}$.

The GMC's at the Galactic center are threaded by a pervasive magnetic
field, whose strength is revealed by the presence of non-thermal
filaments (NTFs) in the diffuse interstellar medium (Morris 2007). The strongly
polarized synchrotron emission from the NTFs indicates that the magnetic field
points along the filaments, whose apparent rigidity when they interact with
molecular clouds and the turbulent interstellar medium suggests field strengths
on the order of a few milligauss (see, e.g., Yusef-Zadeh and Morris 1987).

Confirming evidence for such field strengths in and around the GMC's
is provided by their apparent stability, which does not appear to be
due to the confining pressure of the surrounding medium. The observed
pressure $P_{plasma}$$\sim$$10^{-9.2} \;{\rm erg}\;{\rm cm}^{-3}$
due to the hot plasma between the clouds is an order of magnitude
smaller than that required since the turbulent pressure within them is
$P_{turb}$$\sim$$10^{-8} \;{\rm erg}\;{\rm cm}^{-3}$ (G\"{u}sten and
Philipp 2004). Clouds may instead be bound by their own magnetic fields.
Equating the turbulent and magnetic ($B^2/8\pi$) energy densities gives
field strengths of $\sim 0.5$ mG within the clouds, not too different
from the typical value measured in the NTFs.

We are now also reasonably sure of the magnetic field strength between
the clouds. In the past, the field intensity near the Galactic center
had been uncertain by two orders of magnitude. We've just seen how on
a scale of $\sim$100 pc the NTFs contain field strengths as high as
$\sim$1 mG (see also Yusef-Zadeh and Morris 1987, and Morris
and Yusef-Zadeh 1989), implying a magnetic energy density more than
10,000 times greater than elsewhere in the Galaxy. At the other extreme,
equipartition arguments based on radio observations favor fields of only
$\sim$6 $\mu$G on $\sim$400 pc scales (LaRosa et al. 2005). But a more
careful analysis of the diffuse emission from the central bulge has
revealed a down-break in its non-thermal radio spectrum, attributable
to a transition from bremsstrahlung to synchrotron cooling of the in
situ cosmic-ray electron population. Crocker et al. (2010) have shown
recently that this spectral break requires a field of $\sim$50 $\mu$G
extending over several hundred parsecs, lest the synchrotron-emitting
electrons produce too much $\gamma$-ray emission given existing
constraints (Hunter et al. 1997).

For the purposes of this paper, we will therefore assume an average inter-cloud
magnetic field strength $B \sim\,$30--50$\mu$G throughout the inner several
degrees of the Galaxy, with a corresponding inter-cloud density
$n = 10$ cm$^{-3}$ (again consistent with the limits
placed by Crocker et al. 2010 on the diffuse bremsstrahlung and synchrotron
emissivities), and an associated temperature $T_{plamsa}$$\sim$$5\times 
10^5-5\times 10^6$ K (see, e.g., Belanger et al. 2004, 2006). The
conditions much closer to Sagittarius A* are somewhat different and
appear to be controlled primarily by ongoing stellar wind activity
(Rockefeller et al. 2004). But this is a very small region compared
to the rest of the TeV emitting gas, so we do not expect it to
significantly influence our results.

\section{Calculational Procedure}
If purely turbulent, the magnetic field that permeates the inter-cloud
environment can be treated as a superposition of many
randomly polarized transverse waves which span
a large range of wavelenghts
\begin{equation}
\delta {\bf B} = \sum_\lambda \delta {\bf B'}_\lambda\, {\rm exp}\left[i{2\pi\over\lambda}
(x'-v_At)\right]\,,
\end{equation}
where the primed frame is unique to each wave and $\delta {\bf B'}_\lambda \cdot \hat x' = 0$
(e.g., Giacolone \& Jokipii 1994; Fraschetti \& Melia 2008).
While the exact
nature of this turbulence is not well-constrained, waves are
typically assumed to have energy densities
which follow a power-law spectrum, so that
$(\delta B_\lambda)^2 \sim \lambda^\Gamma$.
As such, a purely turbulent magnetic field
is dominated by the longest
wavelength fluctuations.  Assuming that these fluctuations propagate
at an Alfv\'en speed $v_A \approx \delta B /\sqrt{4\pi\,m_p\,n}$,
a turbulent electric field $\delta \epsilon \sim (v_A/c)\, \delta B$
must also be present as required by
Faraday's law (Fraschetti \& Melia 2008).  Under the most
ideal conditions, the turbulent electric field can energize protons
over a time $\Delta t$ by an amount
\begin{equation}
\Delta E_p \approx e\,\delta\epsilon\,c\,\Delta t \approx e\, \delta B \, v_A\,
\Delta t\;.
\end{equation}
However, such an ideal acceleration can only occur over a length-scale
$\lambda_{max}$ (and hence, a time $\Delta t = \lambda_{max}/c$),
beyond which the process becomes stochastic.

In principle, the kinematics of protons injected in the inter-cloud
environment can be determined by numerically solving the governing
equations of motion (e.g., Wommer et al. 2008).
For the present work, however, the large difference between the
radii of gyration ($\sim 10^{-5}$ pc) of TeV protons in a $50\, \mu$G field
and the size ($\sim 300$ pc) of the inter-cloud region
makes such an approach computationally taxing.
However, the results of such numerical investigations into the kinematics of
$> 10^2$ TeV  protons (which require less computational time) indicate that a
simple, random-walk model can adequately capture the
essential features of particle diffusion and acceleration within the inter-cloud medium.
Specifically, the spatial motion of particles is well approximated by a simple
3-D random walk with a step-size $R_s \sim \lambda_{max}$.  In turn, the
evolution of a particle's energy is well-approximated by a
1-D random-walk in which the particle
gains/loses energy
\begin{equation}
E_s = \chi\,\Delta E_p
 = 5.3 \,{\rm TeV}\, \chi\,\left({\delta B\over 50\, \mu{\rm G}}\right)^2
 \left({R_s\over 1\, {\rm pc}}\right)\;,
 \end{equation}
at each spatial step, with $\chi$ randomly selected between $-0.5$ and $0.5$.

We therefore adopt a Monte Carlo scheme in which $N_p$ particles
are injected randomly (but with uniform probability)
within a radius $R_{acc}$ of the Galactic Center, excluding the regions
occupied by the 14 dominant GMCs in the GC region.
The positions along the plane of the sky of these GMCs are taken from Oka et al. (1998),
with the unknown line-of-sight positions determined via a randomization
process constrained by the requirement that the line-of sight
distribution of these clouds matches
the Galactic plane distribution (see Fig.1 and Table 1 in Wommer et al. 1998).
Particles random-walk through the inter-cloud medium with
a specified step-size $R_s$ until
they either move beyond an escape radius $R_{esc}$  or they
encounter one of the GMCs.  Particles that encounter a GMC are
assumed to $pp$ scatter in the high-density medium, resulting in the
production of neutral pions and subsequent decay photons.
Each particle's contribution to the ensuing gamma-ray emissivity is calculated
using the expression
\begin{equation}
Q_\gamma(E_\gamma) = 2\, c \,n\,\sigma_0\, \int_{E_{\pi_0}^{min}}
dE_{\pi_0}\,{\left[0.67(1-E_{\pi_0}/E_p)^{3.5}+0.5 e^{-18E_{\pi_0}/E_p}\right]
\over E_{\pi_0}\,\sqrt{E_{\pi_0}^2-m_{\pi_0}^2 c^4}}\,
\end{equation}
(e.g., Fatuzzo \& Melia 2003),
where  $\sigma_0 = 32$ mbarns,
$E_{\pi_0}^{min}=E_\gamma +m_{\pi_0}^2 c^4/[4 E_\gamma]$, and
$E_p$ is calculated for each particle using the 1-D random walk
scheme detailed above.  The total emissivity is then found by summing over
the full ensemble of $N_p$ protons.

\begin{figure}
\center{\includegraphics[scale=0.80,angle=0]{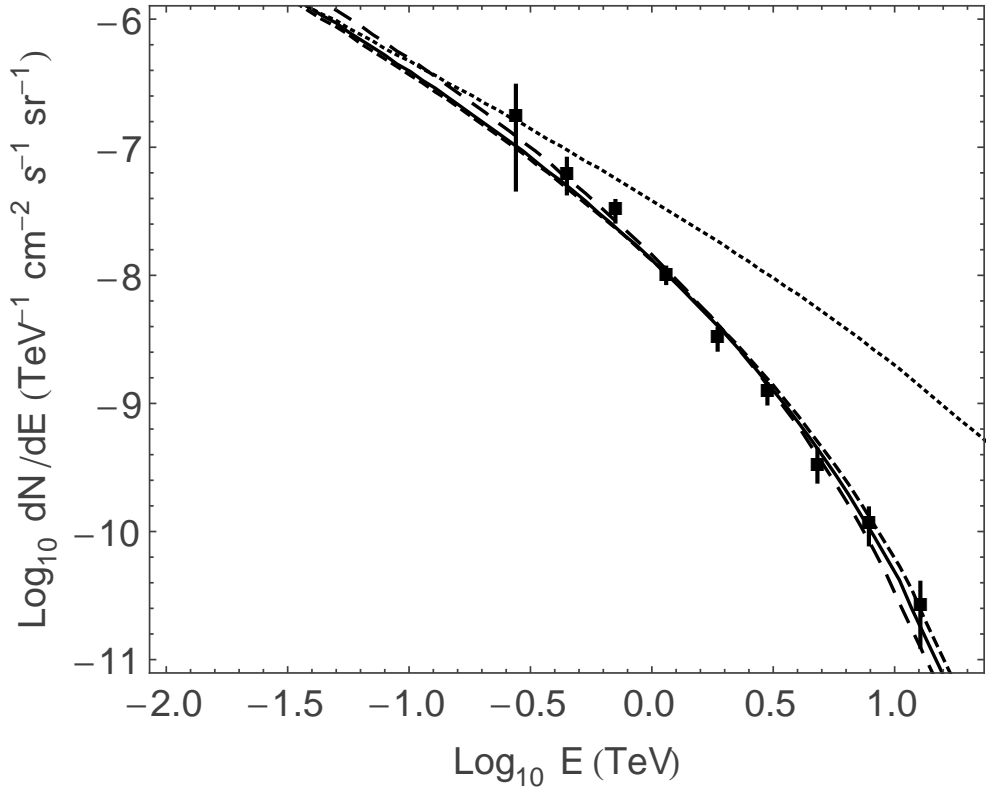}
\vspace{10pt}
\caption{The $\gamma$-ray spectra resulting from the decay of neutral pions
produced by the scattering of protons accelerated diffusively within the
inter-cloud region of the GC and the dense molecular gas
contained within the 14 large GMCs occupying this region.  The curves
show the results of our simple random-walk model of this process for
a purely turbulent inter-cloud magnetic field and four scenarios as defined by
the following parameters:  dotted -- $R_{acc} = 250$ pc, $R_{esc} = 500$ pc,
$\delta B = 50 \,\mu$G, $\eta = 1$; solid --  $R_{acc} = 250$ pc, $R_{esc} = 500$ pc,
$\delta B = 50 \,\mu$G, $\eta = 0.001$; long-dashed --  $R_{acc} = 150$ pc,
$R_{esc} = 250$ pc, $\delta B = 50 \,\mu$G, $\eta = 0.002$; short-dashed --
 $R_{acc} = 250$ pc, $R_{esc} = 500$ pc,
$\delta B = 30 \,\mu$G, $\eta = 0.01$.  The step-size $R_s$ and inter-cloud
density were set to $1$ pc and $10$ cm$^{-3}$, respectively,
for each scenario.  The HESS data are take from Aharonian et al. (2006).}}
\end{figure}

We begin our investigation by calculating the kinematics and resulting
emissivity for $N_p = 10^5$ protons for the model parameters $R_s = 1$ pc,
$R_{acc} = 250$ pc, $R_{esc} = 500$ pc,  and $\delta B = 50 \mu$G.   A comparison
between the (normalized) emissivity (dotted curve) and the HESS data
is presented in Figure 1.  Clearly, this scenario cannot account for the
observations.  We note, however, that the curvature exhibited by the
dotted curve at $> 1$ TeV energy suggests that such a scenario could work
if particles are energized to smaller values when they interact with the
molecular clouds.    We therefore consider the possibility that
particles can only be energized in small, compact regions within
the inter-coud medium, and thereby add an ``efficiency" parameter $\eta$ that
represents the fractional volume of the inter-cloud region within which
particles can gain/lose energy.  For the model parameters adopted above,
a good fit to the data can be obtained with an efficiency of $\eta = 0.001$,
as shown by the solid curve in Figure 1.  While the resulting fit to the data
is quite good, it may be hard to reconcile such a low value of efficiency.
As such, we consider next an inter-cloud medium with the same values of
$\delta B = 50 \,\mu$G and $R_s = 1$ pc, but assume $R_{acc} = 150$ pc and
$R_{esc} = 250$ pc, since a smaller region results in less particle
acceleration.  Indeed,
we find that for these model parameters, good fits to the HESS data
can be obtained with an efficiency of $\eta = 0.002$, as shown
by the long-dashed curve in Figure 1. This efficiency is
still lower than may be reasonably expected.  We next consider
an inter-cloud medium with a magnetic field strength
$\delta B = 30\,\mu$G (but keeping the values of $R_s$,
$R_{acc}$ and $R_{esc}$
unchanged from our initial case).  The efficiency required to obtain
a good fit to the data, as shown by the short-dashed
curve in Figure 1, is now $\eta = 0.01$.

\begin{figure}
\center{\includegraphics[scale=0.80,angle=0]{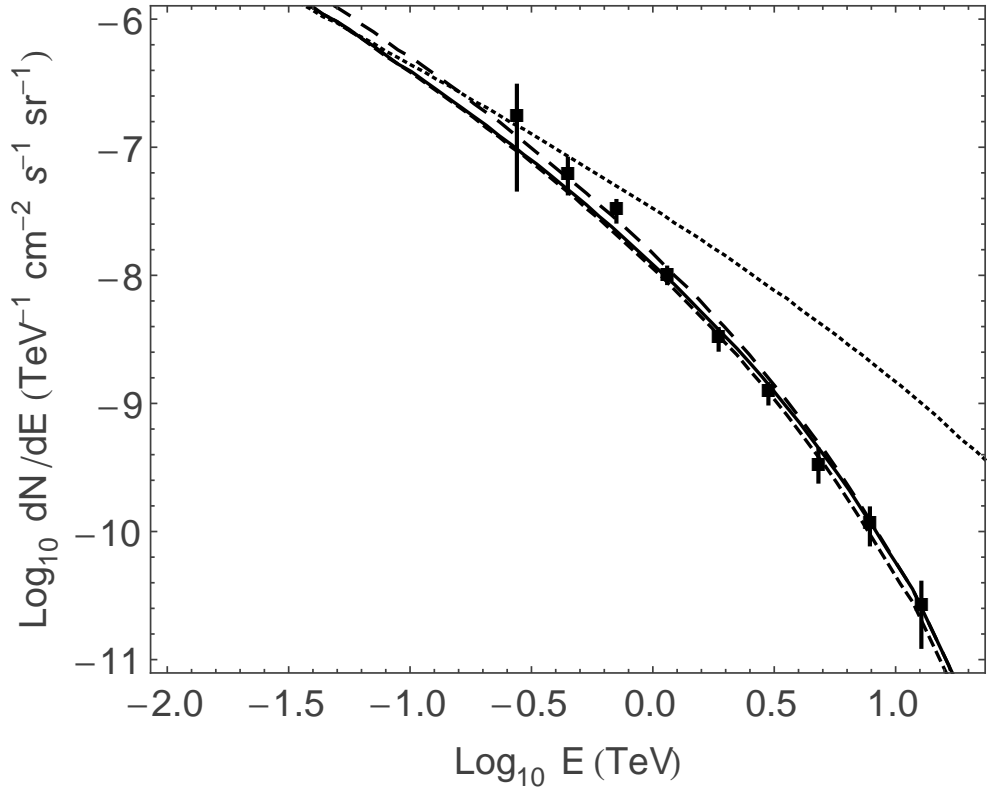}
\vspace{10pt}
\caption{Same as Figure 1, but for the case of
a uniform magnetic field ${\bf B_0}$ directed perpendicular to the Galactic plane
superimposed with a turbulent component $\delta {\bf B}$.  The curves
represent the results of our model for  four scenarios as defined by
the following parameters:  dotted -- $B_0 = 50 \,\mu$G,
$\delta B = 50 \,\mu$G, $\xi = 10$; solid --  $B_0 = 50 \,\mu$G,
$\delta B = 2 \,\mu$G, $\xi = 25$; short-dashed --
 $B_0 = 30 \,\mu$G,
$\delta B = 3 \,\mu$G, $\xi = 10$; long-dashed curve --
$B_0 = 30 \,\mu$G,
$\delta B = 3 \,\mu$G, $\xi = 20$.  The remaining parameters were
set as $R_s = 1$ pc,  $R_{acc} = H_{acc} = 200$ pc,  $R_{esc} = H_{esc} = 
400$ pc  and $n = 10$ cm$^{-3}$
for each scenario.  The HESS data are take from Aharonian et al. (2006).}}
\end{figure}

To understand this strong dependence between the efficiency required to obtain
a good fit to the HESS data and the adopted field strength, let us consider
how {\bf B} and $\eta$ affect the energy distribution of particles
that interact with the molecular clouds. Specifically, the 1-D random
walk in energy leads to a peaked energy distribution with a high-energy tail.
It is this tail of the distribution that produces the
curvature in the ensuing photon emissivity required to fit the HESS data.
Since the energy gained/lost at each step scales as $\delta B^2$, as shown in Equation (3),
the entire distribution shifts accordingly in energy.  That is, doubling
the magnetic field strength shifts the energy distribution upward in energy
by four times.   In addition, the random walk nature of the acceleration
process means that the tail of the energy distribution shifts up in
energy proportionally to the square root of the number of steps
during which the particle gains or loses energy, which is set by the
efficiency.   As such, the energy that characterizes the high-energy
tail (and which is associated
with the HESS data) scales as $E_{p;\,tail} \propto \delta B^2\, \sqrt{\eta}$.
In turn, the efficiency required to fit the HESS data (which requires a
specific value of $E_{p;\,tail}$) scales as $\eta \propto \delta B^{-4}$.
In contrast to this strong dependence on $\delta B$, the direct proportionality
between $\Delta E_p$ and $R_s$ means that our results are not sensitive
to the value of $R_s$ adopted, so long as $R_s << R_{acc}$.

In principle, good fits to the HESS data can be achieved with higher
efficiencies if the inter-cloud magnetic field has a strength significantly
less than $30 \,\mu$G.  Such a scenario for a purely turbulent field appears
ruled out from observations (as noted in \S2). However, it is quite reasonable to expect
that the inter-cloud medium is threaded by a large-scale, static field ${\bf B_0}$ on
which the turbulent magnetic field $\delta {\bf B}$ is superimposed.  In this case,
$\Delta E_p \propto v_A\, \delta B \propto B_0\,\delta B$. To test this scenario,
we have therefore also considered a model in which a uniform
field ${\bf B_0}$ cuts perpendicular to the plane of the galaxy, adopting
a cylindrical acceleration region defined by a radius $R_{acc} = 200$ pc
and a scale height above and below the Galactic plane $H_{acc} = 200$ pc.
Since particles will diffuse preferentially along the direction of the underlying
magnetic field (e.g., Giacolone \& Jokipii 1994), we increase the step-length
along the uniform field direction by a factor $\xi$ (expected to be of order 10
from numerical experiments with higher energy particles).
As before, protons random walk until they either escape from a cylindrical
boundary defined by $R_{esc} = 400$ pc and $H_{esc} = 400$ pc,
or they enter one of the GMCs.   For this case, particles gain/lose energy
\begin{equation}
E_s = \chi\,\Delta E_p
= 0.53 \,{\rm TeV}\, \chi\,\left({B_0\over 50\, \mu{\rm G}}\right)\,
\left({\delta B\over 0.5\, \mu{\rm G}}\right)\,
\left({R_s\over 1\, {\rm pc}}\right)
\end{equation}
during every step where, as before, $\chi$ is randomly chosen between $-0.5$ and 0.5.

Our (normalized) results are illustrated by the dotted curve in Figure 2 for the case
where $B_0 = \delta B = 50 \,\mu$G and $\xi = 10$.  As with the
purely turbulent field, this scenario cannot account for the observations.
We therefore explore  weak-turbulence scenarios, obtaining good fits to the data
for the following model parameters: 1) $B_0 =  50 \,\mu$G, $\delta B = 2 \,\mu$G,
$\xi = 20$ (solid curve); 2) $B_0 =  30 \,\mu$G, $\delta B = 3 \,\mu$G,
$\xi = 10$ (short-dashed curve); and 3) $B_0 =  30 \,\mu$G, $\delta B = 3 \,\mu$G,
$\xi = 20$ (long-dashed curve).  Clearly, a weak turbulence can account
for the observations.  Of course, a strong turbulence scenario could also
account for the observations, but would then require a low acceleration efficiency
$\eta$, as was found for the case of pure turbulence.

For all the cases we have explored here, the $\gamma$-ray spectrum between $100$ MeV 
and $100$ GeV is well-described by a power law of the form $dN/dE \propto E^{-1}$.

We can therefore easily compare our results to the COS-B observations of the  $300$ MeV--$5$ 
GeV emission from the Galactic disk, which includes a possible galactic-center source. 
Integrating the results of our calculated spectrum
normalized to the HESS data (as shown in Figures 1 and 2) over the COS-B energy range, 
we infer a flux of $\sim 3 \times 10^{-11}$ ph cm$^{-2}$ s$^{-1}$, coming from a region
with $\Delta l = 2^0$, $\Delta b = 0.5^0$ centered on the molecular cloud distribution. This 
value is well below the observed total flux $\sim 10^{-7}$ ph cm$^{-2}$ s$^{-1}$ from this region
(Blitz et al. 1985). It would thus appear that pionic decays produced by the
stochastic acceleration of protons in a turbulent field can account for the HESS detected 
TeV emission while at the same time not producing a galactic center $\sim 1$ GeV signal
above the measured COS-B flux.  
We note, however, that $pp$ scattering also produces 
charged pions which quickly decay into secondary leptons. Over time, this leptonic 
population can grown and radiate in the $\gamma$-ray band via synchrotron and Bremmstrahlung 
processes (see, e.g., Fatuzzo \& Melia 2005). Alternatively, the electric fields
associated with this turbulent magnetic field may also accelerate primary electrons.  
Conceivably, both primary and secondary leptons may represent viable sources of
$\sim 1$ GeV emission and ought to be included in a more complete calculation of
the resulting spectra from these particles, which would
require additional model parameters (see, e.g., Rockefeller et al. 2004; Belanger
et al. 2004; Liu et al. 2006; Crocker et al. 2010) and is beyond the scope of the 
present work. We will therefore explore this important issue elsewhere.

\section{Conclusions}
We have shown in this paper that diffusive acceleration of CRs by a
turbulent Alfv\'enic magnetic field spread throughout the inner few
hundred parsecs of the Galaxy can produce an effectively isotropized
distribution of relativistic particles reaching the outer layers of
molecular clouds, where they scatter with other protons to produce
the diffuse TeV glow measured by HESS. The cosmic-ray distribution
produced by this stochastic acceleration results in an excellent
match to the observed TeV spectrum.  Such a scenario, however,
requires a low acceleration efficiency or a weak turbulent field superimposed on
a stronger, large scale static field.

Interestingly, not all of the CRs remain trapped in this region. The long
diffusion length ($\sim 1$ pc) emerging from our simulations ensures
that the cosmic-ray distribution produced in this fashion is isotropized
throughout the molecular gas region where TeV emission occurs. But this
also means that a large fraction of these energetic particles
escape into the rest of the
Galaxy. Calculating the evolution in their distribution as they diffuse
outwards is beyond the scope of this paper. It is clear, however, that
such a simulation ought to be carried out to see if these cosmic rays
contribute noticeably to the distribution reaching Earth. The strength
of this proposal lies in the tightly constrained cosmic-ray population
within the inner bulge, which we infer from the requirements to produce
the observed HESS spectrum. Thus, a detailed simulation of the cosmic-ray
diffusion through the Galactic disk with the initial conditions we infer
from HESS may produce the first evidence that perhaps most of the cosmic
rays we see here are accelerated diffusively in the inner few hundred
parsecs of the Galaxy.

It is natural to wonder whether this process can also occur elsewhere.
Why should it only be evident within the inner few degrees? The answer
is rather straightforward, having to do with the unique physical
conditions we see there. For example, $B$ is much bigger---by at least
a factor of $100$ or maybe $1000$---compared to elsewhere in the disk,
so protons spend much more time in the ``acceleration" zone. Secondly,
both the stellar density and star formation rates are greater there
compared to elsewhere, so the interstellar medium appears to be much
more dynamic. The velocity and density fluctuations in the ISM are
consequently much more significant in the inner bulge than elsewhere,
so although this type of proton acceleration can in principle occur
everywhere, it is much more effective and rapid at the Galactic
center. We are continuing to develop this promising picture of
cosmic-ray acceleration in the Galaxy and will report the results
elsewhere.

\section{Acknowledgments}
This research was partially supported by ONR
grant N00014-09-C-0032 at the University of Arizona, and a Miegunyah
Fellowship at the University of Melbourne. MF acknowledges the support
from the Hauck Foundation at Xavier University. We acknowledge many
helpful discussions with Randy Jokipii, Elizabeth Todd, and Huirong Yan.

\label{lastpage}

\end{document}